\documentclass[12pt]{article}
\usepackage{a4wide}
\usepackage{amsmath,amssymb,amsbsy,enumitem}
\usepackage{latexsym,layout,amsfonts,amssymb,fontenc,xcolor,amsthm,multirow,amsfonts,bm,textcomp,epstopdf}
\usepackage{multirow}
\usepackage{slashbox}
\newsavebox{\uuunit}
\sbox{\uuunit}
    {\setlength{\unitlength}{0.825em}
     \begin{picture}(0.6,0.7)
        \thinlines
        \put(0,0){\line(1,0){0.5}}
        \put(0.15,0){\line(0,1){0.7}}
        \put(0.35,0){\line(0,1){0.8}}
       \multiput(0.3,0.8)(-0.04,-0.02){12}{\rule{0.5pt}{0.5pt}}
     \end {picture}}

\def\2{\frac12}
\def\4{\frac14}

\newcommand{\be}{\begin{equation}}
\newcommand{\ee}{\end{equation}}
\newcommand{\bea}{\begin{eqnarray}}
\newcommand{\eea}{\end{eqnarray}}

\begin{document}

\begin{titlepage}
\begin{center}

\hfill ROM2F/2014/08

\vskip 2cm

{\Large \bf Non-geometric orbifolds and wrapping rules}

\vskip 2cm

{\bf  Gianfranco Pradisi\,$^{1,2}$ and Fabio Riccioni\,$^3$}

\vskip 25pt

{\em $^1$ \hskip -.1truecm Dipartimento di Fisica, Universit\`a di Roma ``Tor Vergata''\\ Via della Ricerca Scientifica 1, 00133 Roma, Italy
 \vskip 15pt }

{\em $^2$ \hskip -.1truecm INFN, Sezione di Roma ``Tor Vergata''\\ Via della Ricerca Scientifica 1, 00133 Roma, Italy
 \vskip 15pt }

{\em $^3$ \hskip -.1truecm
 INFN Sezione di Roma,  Dipartimento di Fisica, Universit\`a di Roma ``La Sapienza'',\\ Piazzale Aldo Moro 2, 00185 Roma, Italy
 \vskip 15pt }

{email addresses: {\tt  Gianfranco.Pradisi@roma2.infn.it},  {\tt Fabio.Riccioni@roma1.infn.it}} \\

\end{center}

\vskip 0.5cm

\begin{center} {\bf ABSTRACT}\\[3ex]
\end{center}

We show that the number of half-supersymmetric $p$-branes in the Type II theories compactified on orbifolds is determined by the wrapping rules recently introduced, provided that one accounts correctly for both geometric and non-geometric T-dual configurations. Starting from the Type II theories compactified on K3, we analyze their toroidal dimensional reductions, showing how the resulting half-supersymmetric $p$-branes satisfy the wrapping rules only by taking into account all the possible higher-dimensional origins. We then consider Type II theories compactified on the orbifold $T^6/(\mathbb{Z}_2 \times \mathbb{Z}_2 )$, whose massless four-dimensional theory is an ${\cal N}=2$ supergravity. Again, the wrapping rules are obeyed only if one includes the complete orbit of the T-duality group, namely either Type IIA or Type IIB theories compactified on either the geometric or the non-geometric T-dual orbifold. Finally, we comment on the interpretation of our results in the framework of the duality between the Heterotic string compactified on ${\rm K3} \times T^2$ and the Type II string compactified on a Calabi-Yau threefold.

\end{titlepage}

\newpage
\setcounter{page}{1}

\tableofcontents

\newpage

\setcounter{page}{1} \numberwithin{equation}{section}

\section{Introduction}

BPS $p$-branes have played a crucial role in many of the developments of String theory \cite{Witten:1995ex, Hull:1994ys, Polchinski:1998rq}.  The fact that the tension of these objects is protected from receiving quantum corrections allows one to rely on their appearance as supergravity solutions to gain information on their properties in the quantum theory. A BPS $p$-brane is electrically charged under a $(p+1)$-form potential, while its dual is the magnetically charged object, whose presence is a manifestation of a ``democracy'' related to the full non-perturbative conjectured quantum symmetry of the theory \cite{Townsend:1996xj}.  Requiring asymptotic flatness of the semiclassical  ``soliton'' solutions at infinity selects $p$-branes with more than two transverse directions, that are charged under the potentials of the  supergravity theory and their magnetic duals.  Nonetheless, in string theory branes are also present with two transverse directions like for instance the D7-brane of Type IIB, with one transverse direction like the D8-brane of the Type IIA, and even with no transverse directions at all, like the D9-brane of Type IIB, that plays a major role in the construction of the Type-I string, determining the background charge of the vacuum \cite{Polchinski:1995mt}.  In $D$ dimensions the branes with two transverse directions are electrically charged with respect to $(D-2)$-form potentials dual to the scalars, while the branes with one and zero transverse directions are coupled to $(D-1)$ and $D$-form potentials not carrying  any propagating degree of freedom and whose existence can be determined only by requiring the closure of gauge and supersymmetry algebras.

In this paper we will study 1/2-BPS branes, preserving the largest possible fraction of supersymmetry. Their complete classification was recently obtained for maximal supergravity theories in any dimensions \cite{Bergshoeff:2011qk, Kleinschmidt:2011vu, Bergshoeff:2012ex}. A crucial ingredient to achieve this result was the identification, as representations of the global symmetry group, of all the potentials of the various maximal theories, including the $(D-1)$-forms and the $D$-forms. Originally, for Type IIA and Type IIB theories the classification was obtained imposing the closure of the supersymmetry algebra \cite{Bergshoeff:2005ac,Bergshoeff:2006qw}.  Later, the result was extended to all dimensions in \cite{Riccioni:2007au,Bergshoeff:2007qi} using the $\text{E}_{11}$ Kac-Moody algebra \cite{West:2001as} and in \cite{deWit:2008ta} using the embedding tensor formalism \cite{Nicolai:2000sc}.  Within the algebraic context, the components coupled to the 1/2-BPS branes are those associated to the longest weights \cite{Bergshoeff:2013sxa}, corresponding to the real roots of the $\text{E}_{11}$ algebra \cite{Kleinschmidt:2011vu}.

The longest weight rule can be formulated in terms of simple light-cone rules for the representations of $\text{SO}(d,d)$ occurring in the decomposition of the global symmetry group as
  \begin{equation}
  G \supset \mathbb{R}^+ \times \text{SO}(d,d)
  \label{decsodd}
  \end{equation}
in $D=10-d$ dimensions,  where  $\mathbb{R}^+$ is the dilaton shift symmetry and $\text{SO}(d,d)$ is the perturbative symmetry whose discrete counterpart is T-duality. Denoting with $i\pm$, $i =1,...,d$ the lightlike directions of $\text{SO}(d,d)$, the longest weights of a given representation correspond to the components satisfying the following light-cone rules \cite{Bergshoeff:2011zk,Bergshoeff:2012ex}\,\footnote{There is also a rule for the spinor representations of $\text{SO}(d,d)$ \cite{Bergshoeff:2011ee} that we ignore because it will not be needed in this paper.}:
\begin{enumerate}[leftmargin=*]
\item\label{r1} for an antisymmetric representation with $n$ indices, one has to select the combinations $i_1 \pm \ i_2 \pm \ ... \ i_n \pm$ with the $i$'s all different. This results into ${d \choose n} \times 2^n$ components;
\item for a mixed-symmetry representation, with Young Tableau made of two columns of length $m$ and $n$, with $m\geq n$, the $m$ indices satisfy the rule \ref{r1} and the $n$ indices must be parallel to $n$ of the $m$ indices. This selects ${d \choose m} \times 2^m \times {m \choose n}$ components.
\end{enumerate}
These rules will be used at various stages in this paper.

Due to the decomposition in eq. \eqref{decsodd}, the $\text{SO}(d,d)$ symmetry does not affect the string dilaton.  As a consequence, the field representations have a definite dilaton weight $\alpha$, a non-positive integer number, and the tension T of the corresponding brane scales like
  \begin{equation}
  {\rm T} \sim g_S^\alpha
  \end{equation}
with respect to the string coupling $g_S$. For instance, the fundamental string has $\alpha=0$, while the D-branes have $\alpha=-1$. In ten dimensions the NS5-brane has $\alpha =-2$, common both to Type IIA and Type IIB theories, while the Type IIB theory possesses a 7-brane with $\alpha=-3$ (the S-dual of the D7-brane) and a 9-brane with $\alpha=-4$ (the S-dual of the D9-brane). Remarkably, the number of branes with $\alpha \geq -3$ that one gets in {\it any}  dimensions using  the light-cone rules can all be reproduced starting from the $p$-branes of the ten-dimensional theories by means of the
``wrapping rules'' \cite{Bergshoeff:2011mh,Bergshoeff:2011ee}
\begin{align}
\alpha=0 \ : \quad  & \begin{cases} {\rm wrapped}   \ \rightarrow\   \ {\rm doubled}\\ {\rm unwrapped} \  \rightarrow \  {\rm undoubled} \ ,\end{cases} \nonumber\\~\nonumber\\
\alpha=-1 \ :\quad & \begin{cases} {\rm wrapped}   \ \rightarrow\   \ {\rm undoubled}\\ {\rm unwrapped} \  \rightarrow \  {\rm undoubled} \ ,\end{cases} \nonumber\\~\label{allwrappingrulesinonego}\\
\alpha=-2 \ :\quad  & \begin{cases} {\rm wrapped}   \ \rightarrow\   \ {\rm undoubled}\\ {\rm unwrapped} \  \rightarrow \  {\rm doubled} \ ,\end{cases} \nonumber\\~\nonumber\\
\alpha=-3 \ :\quad & \begin{cases} {\rm wrapped}   \ \rightarrow\   \ {\rm doubled}\\ {\rm unwrapped} \  \rightarrow \  {\rm doubled} \ , \end{cases}\nonumber
\end{align}
that allow to find the number of branes in a given dimension knowing the number of branes in one dimension higher. Although for the branes with lower values of $\alpha$ there exists no obvious general rule, in any dimension a specific irreducible representation of $\alpha=-4$ space-filling branes carries a number of states obtainable starting from the $\alpha =-4$ 9-brane of Type IIB by means of the additional wrapping rule \cite{Bergshoeff:2012ex}
 \begin{equation}
 \alpha=-4 \ :\quad{\rm wrapped}   \ \rightarrow\   \ {\rm doubled} \quad .
 \end{equation}
The final outcome is that all the branes of the ten-dimensional theories satisfy specific wrapping rules upon torus dimensional reduction.

Among the 1/2-BPS branes of the maximal theories, the ones with two, one and zero transverse directions exhibit the special feature that their BPS conditions are degenerate. Therefore, two or more branes satisfy the same BPS conditions, so that bound states of them can still be 1/2-BPS. This is actually a general feature of all the 1/2-BPS branes of non-maximal supergravity theories. In this paper, when we refer to 1/2-BPS branes, we always consider single-brane states, keeping in mind that one can always construct bound states that still preserve the same amount of supersymmetry. The classification of half-supersymmetric single-brane states in half-maximal theories  was performed in \cite{Bergshoeff:2012jb}, using the fact that the global symmetries of these theories are orthogonal groups and thus still allow for  the application of the light-cone rules. Considering the half-maximal supergravity theories as the low-energy actions of the Heterotic string compactified on tori, the branes with $\alpha=0$ and $\alpha=-2$ are obtained from the fundamental string and the NS5-brane of the ten-dimensional theory using the same wrapping rules as those of the maximal theory.
In particular, on $T^4$, the Heterotic theory is dual to the Type IIA compactified on K3. In the $T^4/\mathbb{Z}_2$ orbifold limit  of K3, the duality was used in \cite{Bergshoeff:2012jb} to show that, again, the same wrapping rules reproduce the branes of the Type IIA theory on K3, if one only takes into account the so-called bulk orbifold cycles. This was then generalized to any K3 orbifold in \cite{Bergshoeff:2013spa}, where it was also shown that the $\alpha=-3$ and $\alpha=-4$ branes are derived using the wrapping rules, by identifying the theory as
the Type IIB on a non-geometric orbifold.

In the first part of this paper we will show that if one considers lower dimensional Type II theories with the same amount of supersymmetry, that are again dual to the Heterotic theory on a torus, the number of branes follow from the wrapping  rules only if one considers together all the possible ways in which the theory can be constructed starting from ten dimensions. In other words, in order to complete the orbit of the T-duality group, one has to consider the torus reduction of all possible six-dimensional theories, {\it i.e.} not only the Type IIA on a geometric orbifold and the T-dual Type IIB on the corresponding non-geometric one, but also the opposite, namely the Type IIB on a geometric orbifold and the T-dual Type IIA on a non-geometric one. The latter has ${\cal N}=(2,0)$ supersymmetry, and its brane classification was obtained in \cite{Bergshoeff:2012jb}.

In general, the wrapping rules encode the information that a given lower-dimensional theory can be considered not only as arising from  either Type IIA or Type IIB, but also from either geometric or non-geometric orbifolds, and only if this information is implemented correctly one obtains  results that are in agreement with applying the light-cone rules on the representations of the T-duality group. In the second part of the paper we will test this feature on theories with less supersymmetry. In \cite{Bergshoeff:2014lxa} the branes of theories with eight supersymmetries have been considered in general. In the case of theories possessing exceptional global symmetries, the classification relies on studying the reality properties of the weights of the representations, but in the particular case of orthogonal global symmetries the light-cone rules again give the correct answer. Viewing these low-energy theories as arising from the Heterotic theory compactified on ${\rm K3} \times T^n$, with $n=1,2,3$, one can then show that the resulting brane classification is again in agreement with the wrapping rules \cite{Bergshoeff:2014lxa}. Here we are going to consider the same theories as arising from Type II reductions, namely four-dimensional theories resulting from Calabi-Yau compactifications. In particular, we will consider the orbifold $T^6/(\mathbb{Z}_2 \times \mathbb{Z}_2 )$. We will show that the wrapping rules exactly hold if one considers all possible T-dual (geometric and non-geometric) orbifolds, provided each configuration is correctly weighted in a way that will be explained in the paper.

It is well known that the Type II theory compactified on a Calabi-Yau manifold is conjectured to be dual to the Heterotic theory compactified on ${\rm K3} \times T^2$ \cite{Aspinwall:1996mn}. For this conjecture to hold, it is necessary that the two theories have the same low-energy effective action, and in particular the same structure of 1/2-BPS branes. We will review some basic facts about the duality, and we will then discuss it in the context of our results.

The plan of the paper is as follows. In section \ref{k3pert2} we discuss the wrapping rules for the  Type II strings on $K3 \times T^n$, for $n=1,2,3$. In section \ref{z2xz2} we consider the $T^6/(\mathbb{Z}_2 \times \mathbb{Z}_2)$ orbifold. In the first subsection, we discuss how T-duality relates geometric and non-geometric orbifolds, while in the second subsection we show how the branes can be counted using the wrapping rules, once the relative weight of the geometric versus non-geometric orbifolds is taken into account. In section 4 we discuss  our results in the context of the duality between the Heterotic string on $K3 \times T^2$ and the Type II string on Calabi-Yau threefolds. Finally, section 5 contains our conclusions.

\section{Type II on ${\rm K3} \times T^n$}\label{k3pert2}

The reductions of the Type IIA and Type IIB theories on K3  give rise to six-dimensional theories possessing ${\cal N}=(1,1)$ and ${\cal N}=(2,0)$ supersymmetry, respectively. More precisely, one obtains a low energy effective action describing   ${\cal N}=(1,1)$ supergravity coupled to 20 vector multiplets in the Type IIA case  and  ${\cal N}=(2,0)$ supergravity coupled to 21 tensor multiplets in the Type IIB case.
In \cite{Bergshoeff:2012jb}, the duality between the Type IIA theory compactified on K3  and the Heterotic theory compactified on $T^4$  was discussed for what concerns the brane counting in the particular case of the $T^4/\mathbb{Z}_2$ orbifold limit of K3. The result of this analysis is that
the wrapping rules satisfied by the maximal theory on the torus can be generalized to the orbifold case, assuming that only the so called ``bulk cycles''  contribute. The same analysis was refined and generalized to any orbifold in \cite{Bergshoeff:2013spa}, where it was also observed that the wrapping rules can be extended to the $\alpha=-3$ and $\alpha=-4$ branes of the Type IIB theory using T-duality, which maps a geometric  orbifold to a non-geometric one.  In \cite{Bergshoeff:2012jb} the wrapping rules  were also applied to the Type IIB theory on a geometric orbifold, reproducing the number of 1/2-BPS branes of the  ${\cal N}=(2,0)$ theory. In this section we want to first review how the wrapping rules give the right numbers of branes in both six-dimensional theories provided that one considers all together geometric and non-geometric orbifolds as an orbit of the T-duality group. Then we want to
consider the dimensional reduction of these theories to five, four and three dimensions, showing again how
the wrapping rules reproduce the right number of branes provided that one considers them as arising from both the ${\cal N}=(1,1)$ and the ${\cal N}=(2,0)$ theories.

We start by reviewing  in detail some of the results of \cite{Bergshoeff:2013spa}. The low-energy theory describing ${\cal N}=(1,1)$ supergravity coupled to 20 vector multiplets possesses a global symmetry $\text{SO}(4,20)\times \mathbb{R}^+$. We denote the $(p+1)$-form fields associated to the $p$-branes as $A_{p+1, A_1 ...A_n B_1 ...B_m}^{(w)}$, where the  $\text{SO}(4,20)$ vector indices denote the irreducible representation whose Young Tableau has two columns, one of length $n$ and one of length $m$, and $w$ is related to the $\mathbb{R}^+$ weight so that the value of $\alpha$ is given by\,\footnote{In the dual Heterotic theory compactified on $T^4$, the dilaton scaling is simply $\alpha_{Het}=-2w$.}
\begin{equation}
\alpha_{6A}=  2w -(p+1) \quad .\label{valueofalphasixdimN=4}
\end{equation}
The fields associated to 1/2-BPS branes are the 1-forms $A_{1,A}^{(0)}$, the 2-forms $A_{2}^{(0)}$ and $A_{2}^{(1)}$, the 3-form $A_{3,A}^{(1)}$, the 4-form $A_{4, A_1 A_2}^{(1)}$, the 5-forms $A_{5,A_1 A_2 A_3}^{(1)}$ and $A_{5,A}^{(2)}$ and the 6-forms $A_{6, A_1 ...A_4}^{(1)}$ and $A_{6, AB}^{(2)}$.
Using eq. \eqref{valueofalphasixdimN=4} and the light-cone rules reviewed in the introduction, one obtains the numbers of 1/2-BPS branes, reported in table \ref{branesD=6N=11}.

\begin{table}[t]
\begin{center}
\begin{tabular}{|c||c|c|c|c|c|}
\hline \rule[-1mm]{0mm}{6mm} \backslashbox{$p$}{$\alpha$} & $\alpha=0$ & $\alpha=-1$ & $\alpha=-2$ & $\alpha=-3$ & $\alpha=-4$ \\
\hline \hline \rule[-1mm]{0mm}{6mm} 0-brane &  & 8& & &  \\
\hline \rule[-1mm]{0mm}{6mm} 1-brane & 1 & & 1 &&   \\
\hline \rule[-1mm]{0mm}{6mm} 2-brane & & 8&&& \\
\hline \rule[-1mm]{0mm}{6mm} 3-brane & & &24& & \\
\hline \rule[-1mm]{0mm}{6mm} 4-brane & & 8&& 32&\\
\hline \rule[-1mm]{0mm}{6mm} 5-brane & & &8& &16\\
\hline
\end{tabular}
\caption{\sl \footnotesize  The  1/2-BPS branes  of the Type II theory compactified to the ${\cal N}=(1,1)$ theory in six dimensions.
\label{branesD=6N=11}}
\end{center}
\end{table}

The branes with $\alpha\geq -3$ can all be derived from the wrapping rules observing that T-duality relates the Type IIA theory on a geometric K3 orbifold to the Type IIB theory on a non-geometric one \cite{Bergshoeff:2013spa}. The non-geometric nature of the IIB orbifold leads to a non-geometric way of counting the cycles along which the branes wrap. As explained in \cite{Bergshoeff:2013spa}, the non-geometric homology seen by the branes in some cases are equivalent to a change of basis in the {\it quantum homology} of K3.  Considering in particular the D-branes, it is well-known that a T-duality along a 1-cycle exchanges a direction transverse to the brane with a direction parallel to the brane. Thus, one is mapping D-branes wrapped on even cycles on the Type IIA side to D-branes wrapped on odd cycles on the Type IIB side. The geometric K3 orbifold only exhibits even cycles, so the non-geometric nature of the Type IIB orbifold is probed by the way the D-branes behave in the T-dual Type IIB, wrapping the ``geometric cycles'' of the T-dual quantum homology.  In other words one obtains the $\alpha=-1$ branes of the six-dimensional theory by either considering the IIA theory, with the D$p$-branes (with $p$ even) wrapped on the six orbifold 2-cycles, wrapped on the whole orbifold or simply unwrapped, or considering the IIB theory, where the D$p$-branes (with $p$ odd) wrap the four non-geometric 1-cycles or their four dual non-geometric 3-cycles. As was explained in \cite{Bergshoeff:2013spa}, the geometric or non-geometric nature of the cycles ``seen'' by the various branes depends actually on the value of $\alpha$: the branes with even $\alpha$  probe the same geometric cycles as in the Type IIA case, while the branes with  $\alpha=-3$ probe non-geometric odd cycles, exactly as the $\alpha=-1$ branes.

Let us first analyze the branes with $\alpha=-3$. In ten dimensions, the only existing brane with such a value of $\alpha$ is the S-dual of the D7-brane of the Type IIB theory. From the wrapping rules, one then gets $1/2 \times 4 \times 2^4 =  32$ 4-branes, where the factor 1/2 accounts for its presence uniquely in the Type IIB theory, 4 is the number of non-geometric 3-cycles and the last factor is related to the wrapping.  The resulting number precisely agrees with the $\alpha=-3$ entry in table \ref{branesD=6N=11}. The $\alpha=-4$ 9-brane of Type IIB, instead, wraps the whole K3, providing $1/2 \times 2^4 = 8$ 5-branes. They account only for 8 of the 16 $\alpha=-4$ branes in table \ref{branesD=6N=11}. Finally, the  $\alpha=-2$ 5-brane present in both Type IIA and Type IIB gives one 1-brane when it fully wraps, and $6\times 2^2 = 24$ 3-branes when it wraps the possible 2-cycles. As observed in \cite{Bergshoeff:2013spa}, when it does not wrap it gives rise to eight 5-branes, while naively one would expect $2^4=16$ branes.  The extra factor 1/2 naturally comes from the fact that the same brane is also present in the ${\cal N}=(2,0)$ theory with the same multiplicity eight. Therefore, the 16 branes predicted by the wrapping rules must be evenly splitted between the two theories, realizing explicitly the $\mathbb{Z}_2$ projection conjectured in \cite{Bergshoeff:2013spa}.

\begin{table}[t]
\begin{center}
\begin{tabular}{|c||c|c|c|c|c|c|}
\hline \rule[-1mm]{0mm}{6mm} \backslashbox{$p$}{$\alpha$} & $\alpha=0$ & $\alpha=-1$ & $\alpha=-2$ & $\alpha=-3$ & $\alpha=-4$ &$\alpha=-5$\\
\hline \rule[-1mm]{0mm}{6mm} 1-brane & 1 & 8& 1 && &  \\
\hline \rule[-1mm]{0mm}{6mm} 3-brane & &8 &24&8 & &\\

\hline \rule[-1mm]{0mm}{6mm} 5-brane & &8 &8& 48&8&8\\
\hline
\end{tabular}
\caption{\sl \footnotesize  The 1/2-BPS branes of the Type II theory compactified to the ${\cal{N}}=(2,0)$ theory in six dimensions.
\label{branesD=6N=20}}
\end{center}
\end{table}

The same analysis can be performed for the ${\cal N}=(2,0)$ theory. The relevant fields, as representations of $\text{SO}(5,21)$, are a 2-form $A_{2, \hat{A}}$, a 4-form $A_{4, \hat{A}_1 \hat{A}_2}$ and a 6-form $A_{6 ,\hat{A} \hat{B}_1 \hat{B}_2}$ \cite{Bergshoeff:2012jb}, where $\hat{A}$'s are indices of $\text{SO}(5,21)$. In order to determine the value of $\alpha$ for each brane, we decompose the global symmetry  as $\text{SO}(5,21) \supset \text{SO}(4,20) \times \text{SO}(1,1)$, where the first factor is the perturbative symmetry ad the second factor is the dilaton scaling. The value of $\alpha$ is then given by
\begin{equation}
\alpha_{6B}= n_+ - n_- - \tfrac{1}{2} (p+1) \quad ,\label{alpharule20theory}
\end{equation}
where $n_+$ and $n_-$ are the number of $1+ = x+t$ and $1-=x-t$ lightlike indices of $\text{SO}(1,1)$ that occur in the decomposition of the representation. As an example, let us consider the 5-branes, corresponding to the 6-form $A_{6 ,\hat{A} \hat{B}_1 \hat{B}_2}$. According to the light-cone rules, at most one of the two $\hat{B}_1 \hat{B}_2$ can be along $1+$, while the $\hat{A}$ index has to be parallel to either $\hat{B}_1$ or $\hat{B}_2$. Denoting with $A$'s the indices of $\text{SO}(4,20)$, this leads to the following possibilities:
 \begin{eqnarray}
  & & A_{6 , 1+ \ 1+ \ A} \quad \ \rightarrow \quad   8 \ {\rm branes}\nonumber \\
  & & A_{6, A \ 1+ \ A} \quad   \ \ \rightarrow \quad 8 \ {\rm branes}\nonumber \\
  & & A_{6, A_1  A_1  A_2} \quad \ \   \rightarrow \quad 2 \times {4 \choose 2} \times 2^2 = 48 \ {\rm branes}\nonumber \\
  & & A_{6, A \ 1- \ A} \quad \ \ \rightarrow \quad  8 \ {\rm branes}\nonumber \\
  & & A_{6, 1- \ 1- \ A} \quad \   \rightarrow \quad 8 \ {\rm branes}\quad .
  \end{eqnarray}
  The number of branes are then determined using the light-cone rules on the $\text{SO}(4,20)$ indices. From  \eqref{alpharule20theory} the value of $\alpha$ of each brane can be determined. The same can be done for the other representations. The final result is summarized in table \ref{branesD=6N=20}.
\begin{table}[t]
\begin{center}
\begin{tabular}{|c|c|c|c|c|c|c|c|}
\hline \rule[-1mm]{0mm}{6mm} brane & field & $0$ &  $-1$ & $-2$& $-3$& $-4$& $-5$      \\
\hline \hline \rule[-3mm]{0mm}{9mm} 0-brane  & $A^{(1)}_{1}$ & 1&  && & &  \\
\rule[-3mm]{0mm}{6mm}  & $A^{(0)}_{1,A}$ & 1& 8 &1& & &  \\
\hline \rule[-3mm]{0mm}{9mm} 1-brane & $A^{(1)}_{2,A}$ & 1&8& 1 & && \\
\rule[-3mm]{0mm}{6mm}  & $A_{2}^{(0)}$ & &  &1& & &  \\
\hline \rule[-3mm]{0mm}{9mm} 2-brane &$A^{(1)}_{3,A_1 A_2  }$ & &8& 24&8&  &  \\
\hline \rule[-3mm]{0mm}{9mm} 3-brane &$A^{(1)}_{4,A_1 A_2 A_3}$ & & & 24 & 32& 24&  \\
\rule[-3mm]{0mm}{6mm}  & $A^{(2)}_{4,A_1 A_2}$ & & 8 &24& 8& &   \\
\hline \rule[-3mm]{0mm}{9mm} 4-brane &$A^{(1)}_{5,A_1 A_2 A_3 A_4}$ & & & & 32& 16& 32 \\
\rule[-3mm]{0mm}{6mm}  & $A^{(2)}_{5,A B_1 B_2}$ & & 8 &8&48 &8 & 8  \\
\hline
\end{tabular}
\caption{\sl \footnotesize  The  1/2-BPS branes  of the Type II theory compactified on ${\rm K3} \times S^1$ for the different values of the dilaton scaling $\alpha$.
\label{branesIIBK3S1}}
\end{center}
\end{table}
It should be underlined that in this case the number of all the branes with $\alpha \geq -4$ can be derived using the wrapping rules. One can view the theory as either a geometric orbifold of the Type IIB or a non-geometric orbifold of the Type IIA,  mapped one to the other by T-duality. The D-branes probe odd bulk  cycles from the Type IIA perspective and even bulk cycles from the Type IIB perspective, both giving the numbers of $\alpha=-1$ branes reported in table \ref{branesD=6N=20}. The $\alpha=-3$ 7-brane, which only exists in the Type IIB, gives rise to 3-branes by wrapping the whole K3 manifold, and to 5-branes by wrapping the 2-cycles. Using the wrapping rules one gets $1/2 \times 2^4= 8$ 3-branes and $1/2 \times 6 \times 2^4= 48$ 5-branes \cite{Bergshoeff:2012jb}. Finally, the branes with even $\alpha$ are the same as in the ${\cal N}=(1,1)$ theory. In particular, the eight $\alpha=-4$ 5-branes are given by applying the  wrapping rules, while the number of $\alpha=-2$ 5-branes, coming from the unwrapped NS5-branes, is halved with respect to the naive calculation due to the
split between the ${\cal N}=(1,1)$ and ${\cal N}=(2,0)$ theories, as mentioned before.

Let us pass now to consider the reduction to five, four and three dimensions. The relevant fields and their representations under the global symmetry group are all given in \cite{Bergshoeff:2012jb} (see, for instance, table 1, eq. (2.25) and eq. (2.27) of that paper for a list in five, four and three dimensions, respectively). For simplicity, we denote with $A,B,..$ the vector indices of the orthogonal global symmetry group in all dimensions, without any risk of confusion because we treat each dimension separately.

In five dimensions the theory possesses a global symmetry $\text{SO}(5,21)\times \mathbb{R}^+$. Denoting the fields as $A_{p+1, A_1 ...A_n B_1 ...B_m}^{(w)}$, with $w$ the $\mathbb{R}^+$ weight and indices as in the non-chiral six-dimensional case, one gets the 1-forms $A_{1,A}^{(0)}$ and $A_{1}^{(1)}$, the 2-forms $A_{2}^{(0)}$ and $A_{2,A}^{(1)}$, the 3-forms $A_{3,A_1 A_2}^{(1)}$, the 4-forms $A_{4,A_1 A_2 A_3}^{(1)}$ and $A_{4,A_1 A_2}^{(2)}$ and finally the 5-forms $A_{4,A_1 A_2 A_3 A_4}^{(1)}$ and $A_{5,A B_1 B_2}^{(2)}$. While in the Heterotic theory the value of $\alpha$ is simply proportional to $w$ being  $\text{SO}(5,21)$ the global symmetry that arises from the five-dimensional torus reduction, in the Type II case one has a perturbative symmetry $\text{SO}(4,20)\times \text{SO}(1,1)$, where the first factor comes from K3 and the second from the circle reduction. Of course, the perturbative $\text{SO}(1,1)$ symmetry is not the one contained in the decomposition $\text{SO}(5,21) \supset \text{SO}(4,20) \times \text{SO}(1,1)$.  It is rather obtained combining the latter with $\mathbb{R}^+$, so that $\alpha$ is given by
\begin{equation}
\alpha_{D=5}= n_+ - n_- - (p+1) +w
\end{equation}
where $n_+$ and $n_-$ are the number of $1+=x+t$ and $1- = x-t$ lightlike indices of the $\text{SO}(1,1)$ inside $\text{SO}(5,21)$. Decomposing the fields and applying the light-cone rules one gets the branes listed in table \ref{branesIIBK3S1} for the different values of $\alpha$.

\begin{table}[t]
\begin{center}
\begin{tabular}{|c|c|c|c|c|c|c|c|c|c|}
\hline \rule[-1mm]{0mm}{6mm} brane & field & $0$ &  $-1$ & $-2$& $-3$& $-4$& $-5$& $-6$& $-7$     \\
\hline \hline \rule[-2mm]{0mm}{7mm} 0-brane & $A_{1,Aa}$ & 4& 16 &4& & & & &  \\
\hline \rule[-2mm]{0mm}{7mm} 1-brane & $A_{2,ab}$ & && 2 & &&&& \\
\rule[-2mm]{0mm}{6mm}  & $A_{2,A_1 A_2}$ & 1&16&26 &16 & 1 & & &  \\
\hline \rule[-2mm]{0mm}{7mm} 2-brane &$A_{3,A_1 A_2 A_3 a}$ & &16& 96&96& 96 & 16& & \\
\hline \rule[-2mm]{0mm}{7mm} 3-brane &$A_{4,A_1 ...A_4 ab}$ & & & 48 & 128& 128& 128 & 48 & \\
\rule[-2mm]{0mm}{6mm}  &$A_{4,A B_1 B_2 B_3}$ & & 16 & 56& 112 & 112 & 112& 56&16\\
\hline
\end{tabular}
\caption{\sl \footnotesize  The  1/2-BPS branes  of the Type II-theory compactified on ${\rm K3} \times T^2$ for the different values of the dilaton scaling $\alpha$.
\label{branesIIBK3T2}}
\end{center}
\end{table}

In the four-dimensional half-maximal theory the global symmetry is $\text{SO}(6,22)\times \text{SL}(2,\mathbb{R})$. The fields are the 1-form $A_{1,Aa}$, the 2-forms $A_{2, A_1 A_2}$ and $A_{2,ab}$, the 3-form $A_{3, A_1 A_2 A_3 a}$ and the 4-forms $A_{4,A_1 ...A_4 ab}$ and $A_{4, A B_1 B_2 B_3}$ \cite{Bergshoeff:2012jb}, where the notation for the $\text{SO}(6,22)$ vector indices is as before, while $a$ denotes an $\text{SL}(2,\mathbb{R})$ doublet and the pair $ab$ is symmetrised.
From the heterotic perspective, $\text{SO}(6,22)$ is again the perturbative Narain symmetry arising from the six-dimensional torus reduction, while $\text{SL}(2,\mathbb{R})$ acts on the axion-dilaton, where the axion is the dual of the NS-NS 2-form. The value of $\alpha$ in the Heterotic theory is related to the number of indices 1 and 2 of the $\text{SL}(2,\mathbb{R})$ according to $\alpha_{Het, D=4} = n_1 - n_2 - (p+1)$ \cite{Bergshoeff:2012jb}.  In the type-II theory, the perturbative symmetry is $\text{SO}(4,20)\times \text{SO}(2,2)$ where the first factor comes from the K3 reduction and the second from the two-torus. There is also an additional $\text{SL}(2,\mathbb{R})$ acting on the axion-dilaton system, as in the Heterotic theory, but in the Type II case the latter $\text{SL}(2,\mathbb{R})$ is contained in the global $\text{SO}(6,22)$ symmetry. The value of $\alpha$ is thus obtained by the decomposition $\text{SO}(6,22) \supset \text{SO}(4,20) \times \text{SO}(2,2)$. In particular, denoting with $n_+$ and $n_-$ the number of $i+ = x_i + t_i$ and $i- = x_i - t_i$ indices in the lightlike directions of the $ \text{SO}(2,2)$, one gets
 \begin{equation}
 \alpha_{D=4} = n_+ - n_- - (p+1) \quad . \label{alpharule4DN=4}
 \end{equation}
It should be observed that the $\text{SL}(2,\mathbb{R})$ indices of the fields listed above do not play any role in determining the dilaton scaling of the branes. Using eq. \eqref{alpharule4DN=4} and the light-cone rules, one obtains the number of branes listed, for the different values of $\alpha$, in table \ref{branesIIBK3T2}.

\begin{table}[t]\footnotesize
\begin{center}
\begin{tabular}{|c|c|c|c|c|c|c|c|c|c|c|c|c|c|}
\hline \rule[-1.5mm]{0mm}{6mm} brane & field & $0$ &  $-1$ & $-2$& $-3$& $-4$& $-5$& $-6$  & $-7$&$-8$ &$-9$&$-10$&$-11$\\
\hline \hline \rule[-1.5mm]{0mm}{6mm} 0-brane & $A_{1,A_1 A_2}$ & 6& 32 &36& 32& 6& &  &&&&& \\
\hline \rule[-1.5mm]{0mm}{6mm} 1-brane & $A_{2,AB}$ & && 4 & &8&& 4&&&&&\\
\rule[-1.5mm]{0mm}{5mm}  & $A_{2,A_1 ...A_4}$ & 1&32&148 &224 & 310 &224 &148 &32 & 1 &&&  \\
\hline \rule[-1.5mm]{0mm}{6mm} 2-brane &$A_{3,A B_1 ...B_5}$ & &32& 296&672&1248 & 1344&1776& 1344& 1248&672&296&32  \\
\hline
\end{tabular}
\caption{\sl \footnotesize  The  1/2-BPS branes  of the Type II-theory compactified on ${\rm K3} \times T^3$ for the different values of the dilaton scaling $\alpha$.
\label{branesIIBK3T3}}
\end{center}
\end{table}

Finally, let us consider the reduction to three dimensions. The three-dimensional theory possesses a global symmetry $\text{SO}(8,24)$, and the fields are the 1-form $A_{1,A_1 A_2}$, the 2-forms $A_{2, AB}$ and $A_{2,A_1 ...A_4}$ and the 3-form $A_{3,A B_1 ...B_5}$ \cite{Bergshoeff:2012jb}. The low-energy action of the Heterotic theory compactified on a seven-torus has a perturbative symmetry $\text{SO}(7,23)$, and the value of $\alpha$ for the different branes is $\alpha_{Het,D=3} = 2(n_+ - n_- -(p+1))$, where $n_+$ and $n_-$ are, again, the number or $1+$ and $1-$ light-cone indices of the $\text{SO}(1,1)$ entering the decomposition $\text{SO}(8,24)\supset \text{SO}(7,23)\times \text{SO}(1,1)$ \cite{Bergshoeff:2012jb}. In the Type II theory compactified on ${\rm K3} \times T^3$, the perturbative global symmetry is $\text{SO}(4,20)\times \text{SO}(3,3)$. To get the value of $\alpha$, also in this case one has to decompose the representations of the fields under $\text{SO}(8,24)\supset \text{SO}(4,20)\times \text{SO}(4,4)$. The end result is
  \begin{equation}
  \alpha_{D=3} = n_+ - n_- -2 (p+1) \quad ,
  \end{equation}
 where $n_+$ and $n_-$ are the number of $i+$ and $i-$ indices (with $i=1,...,4$) of $\text{SO}(4,4)$, respectively. Using this formula together with the  light-cone rules and the fields listed above, one obtains the branes reported in table \ref{branesIIBK3T3}.

We want to show how the wrapping rules are always obeyed in theories with sixteen supercharges, provided the orbits of the T-duality group are correctly taken into account.  We limit ourselves to list the results for all the branes with $\alpha\geq -3$ in table \ref{wrappingrulesN=4}. In going from 6 to 5 dimensions, the wrapping rules have to be applied as follows: one has to consider only the contribution from either the ${\cal N}=(1,1)$ or the ${\cal N}=(2,0)$ theory if a  brane does not double (so, consistently, one gets the same number from both theories), or one has to sum the two contributions if it doubles. This precisely mimics what happens in the maximal theory going from 10 to 9 dimensions. The reader can appreciate that all the numbers are precisely reproduced by the wrapping rules.  We want to stress that the space-filling branes with $\alpha=-4$ always arise in reducible representations.  As a result, if one decomposes the numbers of such branes appearing in tables \ref{branesIIBK3S1}, \ref{branesIIBK3T2} and \ref{branesIIBK3T3} in terms of the number of branes corresponding to each irreducible representation, one would find, in any dimension, the presence of one representation reproducing the numbers in agreement with the $\alpha=-4$ wrapping rule. We have not included these branes in table \ref{wrappingrulesN=4}, but the consistency checks can be easily worked out.

\begin{table}[t]\scriptsize
\begin{center}
\begin{tabular}{|c||c|c|c|c||c|c|c|c||c|c|c|c||c|c|c|c|}
\hline \rule[-2mm]{0mm}{6mm} $\alpha$ &  \multicolumn{4}{|c||}{$\alpha=0$} &     \multicolumn{4}{|c||}{$\alpha=-1$} & \multicolumn{4}{|c||}{$\alpha=-2$} & \multicolumn{4}{|c|}{$\alpha=-3$}  \\
 \hline \rule[-1mm]{0mm}{6mm} \backslashbox{$p$}{$D$} & 6A/6B &5 &4 &3 & 6A/6B& 5& 4&3& 6A/6B& 5&4 &3& 6A/6B & 5 & 4&3 \\
\hline \hline \rule[-2mm]{0mm}{6mm} 0 &  &2 &4 &6& 8/0 &8 &16 &32& & 1& 4 &36 & & &&32\\
\hline \rule[-2mm]{0mm}{6mm} 1 & 1/1 & 1&1&1 & 0/8 &8 & 16 &32&1/1 &2 & 28&152 & & &16&224\\
\hline \rule[-2mm]{0mm}{6mm} 2 & & && & 8/0 & 8&16&32& & 24& 96 &296& & 8& 96 &672\\
\hline \rule[-2mm]{0mm}{6mm} 3 & & & && 0/8 & 8&16& & 24/24 & 48 & 104 &&0/8 & 40& 240&\\
\hline \rule[-2mm]{0mm}{6mm} 4 & & & && 8/0 & 8 & & & & 8 &  & &32/0& 80 & & \\
\hline \rule[-2mm]{0mm}{6mm} 5 & & & && 0/8 &  & & & 8/8&  &  && 0/48&  & & \\
\hline
\end{tabular}
\caption{\sl \footnotesize  Table showing that the branes with $\alpha\geq -3$ of the type-II theories with sixteen supersymmetries in five, four and three dimensions are derived by the wrapping rules starting from the branes of the two six-dimensional theories. We denote the six-dimensional (1,1) and (2,0) theories as 6A and 6B to emphasise the similarity with the ten-dimensional maximal case.
\label{wrappingrulesN=4}}
\end{center}
\end{table}

The outcome of this section is as follows.  In order to reproduce the number of single-brane states of the Type II theory compactified on ${\rm K3} \times T^n$ using the wrapping rules, one has to consider together all the possible theories related by T-duality that give rise, after reduction, to the same theory.
In the next section we move to consider the four-dimensional Type II theory compactified on a six-dimensional $T^6/(\mathbb{Z}_2 \times \mathbb{Z}_2 )$  orbifold.  In this case T-duality merges geometric and non-geometric compactifications in a more intricate way.  Starting from a geometric orbifold and performing all possible T-dualities, one can count the ratio of geometric versus non-geometric orbifolds.  We shall show that, again, one reproduces the number of $\alpha\geq -3$ branes using the wrapping rules, if and only if one takes correctly into account the orbit of the T-duality group and the relative weight of  geometric versus non-geometric configurations.

\section{Type II on the orbifold $T^6/(\mathbb{Z}_2 \times \mathbb{Z}_2 )$}\label{z2xz2}

As anticipated, in this section we want to show that the wrapping rules predict the number of branes with $\alpha\geq -3$ in the Type II theory compactified to four dimensions on the orbifold $T^6/(\mathbb{Z}_2 \times \mathbb{Z}_2 )$. A crucial ingredient is the observation that T-duality relates geometric and non-geometric orbifolds.  In particular, we are going to show that the wrapping rules work perfectly after taking into account the ratio of geometric versus  non-geometric orbifolds within the orbit of the T-duality group. The relation between T-duality and non-geometric orbifolds is discussed in section \ref{Tduanngeo}, while in section \ref{wrultypeii} the branes of the theory for different values of $\alpha$ are determined, together with the way in which the wrapping rules are at work.

The branes of theories with ${\cal N}=2$ supersymmetry were considered in general in \cite{Bergshoeff:2014lxa}.  In the same paper, it was also shown that the  interpretation in terms of the Heterotic String compactified on ${\rm K3} \times T^n$ allows to reproduce the number of branes using the wrapping rules. On the other hand, it has been conjectured \cite{Kachru:1995wm, Ferrara:1995yx} that the Heterotic String compactified on ${\rm K3} \times T^2$ and the Type II theory compactified on a Calabi-Yau threefold are dual: we shall discuss in section \ref{commduality} some aspects of the relation between the duality and our results.

\subsection{T-duality and non-geometric orbifolds}\label{Tduanngeo}

Type II strings compactified on a Calabi-Yau threefold of Hodge numbers $(h_{11}, h_{12})$ are characterized by massless spectra with ${\cal N}=2$ supersymmetry, due to the $SU(3)$ holonomy of the manifold \cite{Polchinski:1998rq}.  The universal (bosonic) sector contains the graviton, the axion-dilaton field and the $2 (h_{11} + h_{22})$ real scalars from the fluctuations of the ten-dimensional metric and of the Kalb-Ramond field.  Adding the contributions of the R-R sectors, one gets for the Type IIA theory an ${\cal N}=2$ supergravity coupled to the universal hypermultiplet (containing the axion-dilaton), $h_{12}$ additional hypermultiplets and $h_{11}$ vector multiplets. The moduli space is a product
\be
{\cal M}={\cal M}_V  \times {\cal M}_H \
\label{modulispace}
\ee
of the special K\"ahler manifold ${\cal M}_V$ parameterized by the scalars of the vector multiplets and the
quaternionic manifold ${\cal M}_H$ parameterized by the scalars of the hypermultiplets \cite{de Wit:1984px, Aspinwall:2000fd}.
In the Type IIB string the situation is similar, since one gets $h_{11}$ additional hypermultiplets and $h_{12}$ vector multiplets, with a corresponding moduli space of the same form as in eq. (\ref{modulispace}), with the dimensions of the two spaces ${\cal M}_V$ and ${\cal M}_H$ interchanged with respect to the Type IIA string.  Calabi-Yau manifolds with interchanged Hodge numbers, if they exist, form mirror pairs.  As a result, the Type IIA string compactified on a Calabi Yau manifold $X$ exhibits the same moduli space as the Type IIB string compactified on the mirror Calabi-Yau manifold $\tilde X$.

As done in \cite{Bergshoeff:2013spa} for six-dimensional models, we would like to verify the classifications of single-brane half-BPS states obtained in terms of the T-duality groups in an explicit string compactification, where possibly the geometric properties, the moduli space and the action of  T-duality on the compactification manifold be under control.  The simplest of such models is the $T^6/(\mathbb{Z}_2 \times \mathbb{Z}_2 )$ orbifold of the Type II in its full-fledged T-dual orbit.  In order to describe it, a necessary ingredient is the related geometry. Starting with a six-torus of a factorized $T^2\times T^2\times T^2$ form, the orbifold group action is generated by
\begin{align}
&g : \ (z_1, z_2, z_3) \rightarrow (z_1, - z_2, - z_3) \  ,
&h : \ (z_1, z_2, z_3) \rightarrow (-z_1, - z_2,  z_3) \  .
\end{align}
As a consequence, three twisted sectors are present, containing  $16$ fixed tori to be identified with the
untouched two-tori tensorized with the $16$ four-dimensional fixed-points of each of the group elements.  Properties of the (integer) Homology group are encoded in the Hodge diamond.  It can be easily deduced by the Hodge diamond of the covering torus
\be
\begin{array}{ccccccc} &&&h_{00}\\ &&h_{10}&&h_{01}\\ &h_{20}&&h_{11}&&h_{02}\\ h_{30}&&h_{21}&&h_{12}&&h_{33}\\
&h_{31}&&h_{22}&&h_{13}\\
&&h_{32}&&h_{23} \\ &&&h_{33} \end{array}=
\begin{array}{ccccccc} &&&1\\ &&3&&3\\ &3&&9&&3\\ 1&&9&&9&&1\\ &3&&9&&3\\
&&3&&3 \\ &&&1 \end{array} \ ,\label{torushodgediamond}
\ee
considering the invariant (bulk) cycles and the exceptional divisors, corresponding to the fixed tori.  A small resolution of the latter singularities will
provide the corresponding smooth manifold. To find the normalized integer (co)homology, as usual, one has to introduce the ``fractional cycles'' and to carefully choose a corresponding basis.  As a result, the ``bulk'' invariant cycles, connected to the untwisted sector, are given by
\be
\begin{array}{ccccccc} &&&h_{00}\\ &&h_{10}&&h_{01}\\ &h_{20}&&h_{11}&&h_{02}\\ h_{30}&&h_{21}&&h_{12}&&h_{33}\\
&h_{31}&&h_{22}&&h_{13}\\
&&h_{32}&&h_{23} \\ &&&h_{33} \end{array}=
\begin{array}{ccccccc} &&&1\\ &&0&&0\\ &0&&3&&0\\ 1&&3&&3&&1\\ &0&&3&&0\\
&&0&&0 \\ &&&1 \end{array} \ .\label{untwhodgediamond}
\ee
In addition, there are 16 fixed tori in each of the $3$ twisted sectors.  The corresponding minimal resolution of the $\mathbb{A}_1$ singularities, give rise to a contribution $16$ to both the non-trivial Hodge numbers.  However, the group invariant part sets only one of the contributions.  The final geometric Calabi-Yau manifold is characterized by the homology
\be
\begin{array}{ccccccc} &&&h_{00}\\ &&h_{10}&&h_{01}\\ &h_{20}&&h_{11}&&h_{02}\\ h_{30}&&h_{21}&&h_{12}&&h_{33}\\
&h_{31}&&h_{22}&&h_{13}\\
&&h_{32}&&h_{23} \\ &&&h_{33} \end{array}=
\begin{array}{ccccccc} &&&1\\ &&0&&0\\ &0&&3+48&&0\\ 1&&3&&3&&1\\ &0&&3+48&&0\\
&&0&&0 \\ &&&1 \end{array} \ .\label{orbhodgediamond}
\ee
It should be stressed that the light-cone rules select exactly those states linked to the untwisted "bulk" part of the
cycles.  The conformal field theory allows for another solution of the same theory, connected to the
previous one by the so called ``discrete torsion'' \cite{Vafa:1994rv}.  Indeed, the modular invariant allows a discrete deformation corresponding to a different
combination of independent modular orbits.  The result in this simple case does correspond to a compactification of the Type IIA string on the
mirror Calabi-Yau, related to the previous one by the exchange of the Hodge numbers.  It should be noted, however, that the discrete torsion
selects different BPS conditions.  For instance, it is well known by the analysis of D-branes in orientifolds of the Type IIB theory \cite{Angelantonj:2002ct} that it forces the introduction of ``exotic'' orientifold planes, with reversed tension and RR charge, and of the corresponding anti-D-branes.  Namely, the $1/2$-BPS states surviving the orientifold projection in the D9 sector are orthogonal to the ones in the D5 sector.  As a consequence, the corresponding orientifolds exhibit the ``brane supersymmetry breaking'' phenomenon \cite{Antoniadis:1999xk}, with a resulting four dimensional theory no longer supersymmetric.  Another way to understand the same issue is related to the fact that Type IIA with discrete torsion is equivalent to the Type IIB without discrete torsion compactified on the same Calabi-Yau.  Thus, in this very simple context, discrete torsion is mirror symmetry.

We should stress that, as shown in the paper \cite{Bergshoeff:2013spa}, we need to describe a certain compactification in the full-fledged T-duality setting.
It means that we must understand both the $A$- and $B$- type of branes present in a theory and in the mirror symmetric.  The $T^6/(\mathbb{Z}_2 \times \mathbb{Z}_2 )$ orbifold exhibits notably this property.  Indeed, it is very easy to describe Type IIA on the geometric orbifold and at the same time Type IIB on the T-dualized orbifolds: some of them are non-geometric asymmetric orbifolds, others still correspond to geometric compactifications.  Of course, on a generic Calabi-Yau this would be much more difficult to understand, even though probably more general examples can be given.  Like in the case of the orbifold limit of K3, what we have to classify are exactly the relative weights between the different T-dual descriptions.  One efficient way to understand what is the action of T-dualities is trying to identify the cycle in the non-geometric case, namely to describe the T-duality action on the homology.
After the orbifold projection, indeed, we saw that the Hodge diamond restricted by the light-cone rules contains one $0$-cycle, one $6$-cycle, three $2$-cycles, three $4$-cycles and eight $3$-cycles.  They correspond to the supersymmetric cycles for the corresponding D-branes.  Now, one can perform a certain number of T-dualities in directions that are parallel or normal to the cycles of a certain homology basis.  The surviving ``non geometric'' cycles\footnote{It means that they are ``geometric'' in terms of the T-dual coordinates.} are now the invariants under the combined action of the orbifold group and an involution, corresponding exactly to the T-duality.  The resulting homology changes.  For instance, with a single T-duality the ``non geometric'' homology
consists of a vanishing number of $0$- and $6$-cycles, two $1$- and $5$-cycles and four $2$-, $3$- and $4$-cycles.

Let us analyze what are the resulting configurations, taking into account that an odd number of T-dualities maps Type IIA in Type IIB and vice-versa, while an even number of them does not change the theory, being equivalent to a redefinition of some moduli. Starting, for instance, from a Type IIA geometric configuration, one or five T-dualities result into six different non-geometric configurations of the Type IIB.  There are 15 options to perform a pair of T-dualities. The three related to T-dualities along two directions of the same two-torus give rise to three non-geometric Type IIA configurations, while the remaing twelve yield as many Type IIA geometric configurations. T-dualizing the complementary directions, the same result is obtained with four T-dualities.  The options for a T-duality along three directions are 20.  Eight of them result in geometric Type IIB, recognizable as compactifications on the mirror symmetric manifold \cite{Strominger:1996it}, while twelve correspond to non geometric compactifications. Finally, six T-dualities are just redefinition of the Type IIA orbifold radii.  As a result, starting from a Type IIA geometric configuration and acting with T-dualities one gets eight geometric Type IIA and eight geometric Type IIB models, together with 24 non-geometric Type IIA and 24 non geometric Type IIB models.  It should be noticed that the same orbits are obtained starting from a geometric Type IIB compactification. The net result is that, by considering all the configurations as different components of a {\it unique} theory invariant under T-duality, one may appreciate how the weight of geometric versus non-geometric configurations is exactly 16:48, namely 1:3.  As we will see in the next section, this ratio corresponds exactly to what is needed to verify how the wrapping rules are perfectly at work if one considers in a correct way the ``stringy geometry'' of configurations.  Of course, it would be very interesting to extend the counting to more complicated realizations of Calabi-Yau manifolds than the  $T^6/(\mathbb{Z}_2 \times \mathbb{Z}_2 )$ orbifold.

\subsection{Wrapping rules of Type II on  $T^6/(\mathbb{Z}_2 \times \mathbb{Z}_2 )$}\label{wrultypeii}

The analysis of the previous subsection allows us to count properly the geometric and non-geometric cycles that each brane of the Type IIA and Type IIB ten-dimensional theories  probe upon reduction on the $T^6/(\mathbb{Z}_2 \times \mathbb{Z}_2 )$ orbifold. The aim of this subsection is to show that the number of branes with $\alpha\geq - 3$ in the resulting four-dimensional theory is reproduced by the wrapping rules provided that the mentioned counting of cycles is taken into account. We first review the analysis of \cite{Bergshoeff:2014lxa} that classifies the 1/2-supersymmetric branes in symmetric four-dimensional ungauged theories with ${\cal N}=2$ supersymmetry, then we determine the value of $\alpha$ for each brane in the Type II theory. Finally, we discuss the wrapping rules.

The classification of branes in theories with eight supersymmetries performed in \cite{Bergshoeff:2014lxa} is based on the reality properties of the weights of the representations to which the brane charges belong. In particular, one can consider the model in which the global symmetry of the low-energy action is $\text{SO}(2, n_V -1 ) \times \text{SL}(2,\mathbb{R}) \times \text{SO}(4,n_H )$, where $n_V$ and $n_H$ are the number of vector multiplets and hypermultiplets, respectively.  The aforementioned reality properties of the weights select the components of the charges that satisfy the light-cone rules.
Denoting with $A,B,...$ the vector indices of $\text{SO}(2, n_V -1 )$, with $a,b,...$ the doublet indices of $\text{SL}(2,\mathbb{R})$ and with $M,N,...$ the vector indices of $\text{SO}(4, n_H )$, the fields associated to 1/2-BPS branes are the 1-forms $A_{1,Aa}$, the 2-forms $A_{2,ab}$, $A_{2,A_1 A_2}$ and $A_{2, M_1 M_2}$, the 3-forms $A_{3, M_1 M_2 Aa}$ and the 4-forms $A_{4, M_1 M_2 A_1 A_2 ab}$, $A_{4, MN_1 N_2 N_3}$ and $A_{4, AB M_1 M_2}$ \cite{Bergshoeff:2014lxa}. Using the light-cone rules, one can then derive the number of 1/2-BPS branes (see tables 5 and 10 of \cite{Bergshoeff:2014lxa}).

In the Heterotic theory, this model can be thought of as a compactification on ${\rm K3} \times T^2$, where
$\text{SO}(4, n_H )$ is the global symmetry of the moduli from the K3 reduction and $\text{SO}(2, n_V -1 )$ comes from the torus reduction, assuming a phase where the surviving gauge group is broken to its maximal abelian subgroup. Finally, the string dilaton together with the dual of the NS-NS 2-form parametrize the coset $\text{SL}(2,\mathbb{R})/\text{SO}(2)$.  Since $\text{SO}(4, n_H )$ and $\text{SO}(2, n_V -1 )$ are both perturbative symmetries, the value of $\alpha$ of the various branes can only depend on the rank of the form and on the $\text{SL}(2,\mathbb{R})$ indices. The precise relation is $\alpha  = n_1 - n_2 - (p+1)$ for each $p$-brane  \cite{Bergshoeff:2014lxa}, where $n_1$ and $n_2$ are the numbers of up and down indices of $\text{SL}(2,\mathbb{R})$. The resulting number of branes is reported in table \ref{braneshetK3T2}. As shown in \cite{Bergshoeff:2014lxa}, the columns with $\alpha=0$ and $\alpha=-2$ can also be derived applying the wrapping rules.

\begin{table}[t]
\begin{center}
\begin{tabular}{|c|c|c|c|c|c|}
\hline \rule[-1mm]{0mm}{6mm} brane & field & $\alpha=0$ &  $\alpha=-2$ & $\alpha=-4$& $\alpha=-6$ \\
\hline \hline \rule[-2mm]{0mm}{7mm} 0-brane & $A_{1,Aa}$ & 4& 4 &&   \\
\hline \rule[-1mm]{0mm}{6mm} 1-brane & $A_{2,ab}$ &1 &&1 & \\
\rule[-1mm]{0mm}{6mm}  & $A_{2,A_1 A_2}$ & &4& & \\
\rule[-2mm]{0mm}{7mm}  & $A_{2,M_1 M_2}$ & &24& & \\
\hline \rule[-2mm]{0mm}{7mm} 2-brane &$A_{3,M_1 M_2 A a}$ & &96& 96&\\
\hline \rule[-1mm]{0mm}{6mm} 3-brane &$A_{4,M_1 M_2 A_1 A_2 ab}$ & &96& &96\\
\rule[-1mm]{0mm}{6mm}  &$A_{4,M N_1 N_2 N_3}$ & &&96 &\\
\rule[-2mm]{0mm}{7mm}  &$A_{4,A B M_1 M_2}$ & &&96 &\\
\hline
\end{tabular}
\caption{\sl \footnotesize  The  1/2-BPS branes  of the Heterotic theory on ${\rm K3} \times T^2$.
\label{braneshetK3T2}}
\end{center}
\end{table}

We now want to consider the same model from the Type II perspective. In particular, we restrict ourselves to the analysis of the Type II theory compactified on the orbifold $T^6/(\mathbb{Z}_2 \times \mathbb{Z}_2 )$ corresponding, as seen in section \ref{Tduanngeo}, to a Calabi- Yau compactification with ($h_{11}=51, h_{12}=3$).  As it is well-known, for a generic Calabi-Yau compactification one has $n_V = h_{11}$ and $n_H = h_{21} +1 $ in the Type IIA case and $n_V = h_{21}$ and $n_H = h_{11} +1 $ in the Type IIB case. Since in Type II strings the dilaton belongs to a hypermultiplet, the symmetry $\text{SO}(2, n_V )\times \text{SL}(2,\mathbb{R})$ of the vector-multiplet sector is perturbative. The symmetry of the hypermultiplet sector $\text{SO}(4, n_H )$, instead, is broken at the perturbative level to $\text{SO}(2, n_H -2 ) \times \text{SO}(2,2 )$, where $\text{SO}(2,2)$ is isomorphic to $\text{SL}(2,\mathbb{R})\times\text{SL}(2,\mathbb{R})$. One of the two $\text{SL}(2,\mathbb{R})$ can be identified with the symmetry group that transforms the axion-dilaton complex scalar.  As a consequence, the value of $\alpha$ is only a function of the rank of the form and of the numbers $n_+$ and $n_-$ of light-cone indices of $\text{SO}(2,2)$. In particular, one obtains
  \begin{equation}
  \alpha = n_+ - n_- - (p+1) \quad .\label{alphaorbifoldT6Z2Z2}
  \end{equation}
This formula implies that all the $p$-branes whose  charges do not carry indices in the hypermultiplet sector have a value of $\alpha$ which is simply $-(p+1)$. The branes associated to the field $A_{2, M_1 M_2}$, corresponding to defect 1-branes magnetically charged under the hypermultiplet scalars, split as
 \begin{eqnarray}
  & & A_{2 , 1+ \ 2+ } \quad \ \rightarrow \quad   1 \ {\rm brane}\nonumber \\
  & & A_{2,  i+ \ m} \quad   \ \ \rightarrow \quad 2 \times 4 =8 \ {\rm branes}\nonumber \\
  & & A_{2, m_1 m_2} \quad  \ \   \rightarrow \quad {2 \choose 2} \times 2^2 = 4 \ {\rm branes}\nonumber \\
  & & A_{2, i+ \ j-} \quad \ \ \rightarrow \quad  2 \ {\rm branes}\nonumber \\
  & & A_{2, i-  \ m} \quad \  \ \  \rightarrow \quad 2\times 4 =8 \ {\rm branes}\nonumber \\
  & & A_{2, 1- \ 2-} \quad \ \  \rightarrow \quad 1 \ {\rm brane} \quad ,  \label{fourdimdecomplightcone}
  \end{eqnarray}
where $i\pm$ ($i =1,2$) are the light-cone directions of $\text{SO}(2,2)$, $m$'s take values along the light-cone directions of $\text{SO}(2, n_H -2 )$ and the number of corresponding branes has been derived using the light-cone rules. The same decomposition can be performed for all the other fields, {\it i.e.} the 3-form $A_{3, M_1 M_2 Aa}$ and the 4-forms $A_{4, M_1 M_2 A_1 A_2 ab}$, $A_{4,M N_1 N_2 N_3}$ and $A_{4, AB M_1 M_2}$. The value of $\alpha$ of each brane is determined using eq. \eqref{alphaorbifoldT6Z2Z2}.  The resulting numbers are listed in table \ref{branesIIBT6Z2Z2}.

\begin{table}[t]
\begin{center}
\begin{tabular}{|c|c|c|c|c|c|c|c|c|c|}
\hline \rule[-1mm]{0mm}{6mm} brane & field & $0$ &  $-1$ & $-2$& $-3$& $-4$& $-5$& $-6$& $-7$     \\
\hline \hline \rule[-2mm]{0mm}{7mm} 0-brane & $A_{1,Aa}$ & & 8 && & & & &  \\
\hline \rule[-1mm]{0mm}{6mm} 1-brane & $A_{2,ab}$ & && 2 & &&&& \\
\rule[-1mm]{0mm}{6mm}  & $A_{2,A_1 A_2}$ & &&4 & &  & & &  \\
\rule[-2mm]{0mm}{7mm}  & $A_{2,M_1 M_2}$ &1 &8&6 &8 & 1 & & &  \\
\hline \rule[-2mm]{0mm}{7mm} 2-brane &$A_{3,M_1 M_2 A a}$ & &8& 64&48& 64 & 8& & \\
\hline \rule[-1mm]{0mm}{6mm} 3-brane &$A_{4,M_1 M_2 A_1 A_2 ab}$ & & &8& 64&48& 64 & 8&  \\
\rule[-1mm]{0mm}{6mm}  &$A_{4,M N_1 N_2 N_3}$ & & 8 & 12& 24 & 8 & 24& 12&8\\
\rule[-2mm]{0mm}{7mm}  &$A_{4,AB M_1 M_2}$ & & & 4 & 32& 24 & 32 & 4 &\\
\hline
\end{tabular}
\caption{\sl \footnotesize  The  1/2-BPS branes  of the Type II-theory compactified on the orbifold $T^6/(\mathbb{Z}_2 \times \mathbb{Z}_2 )$ for the different values of $\alpha$.}
\label{branesIIBT6Z2Z2}
\end{center}
\end{table}

Let us show how the number of all the branes with $\alpha \geq -3$ in table \ref{branesIIBT6Z2Z2} are derived using the wrapping rules by the weighted inclusion of geometric and non geometric contributions.  It is useful to recall that branes with even $\alpha$ probe always geometric cycles, while branes with odd $\alpha$ probe the non-geometric cycles in the corresponding non-geometric orbifolds.
\begin{itemize}[leftmargin=*]
\item $\alpha=0$ branes. The ten-dimensional fundamental string can never wrap, so it never doubles. The only $\alpha=0$ brane in four dimensions is the unique fundamental string.
\item $\alpha=-1$ branes. The Type IIA theory in ten dimensions contains D$p$-branes, with $p$ even. The geometric orbifold in four dimensions exhibits eight 0-branes, one corresponding to the unwrapped D0-brane, three from the D2-brane wrapped on 2-cycles, three from the D4-brane on 4-cycles and one from the D6 on the whole orbifold. Similarly, there are eight 1-branes, all coming from the D4-brane wrapped on 3-cycles. In the same way, more eight 2-branes and eight 3-branes can be obtained. The non-geometric orbifold contains the same number of branes, but obtained in a different fashion. Indeed, four out of eight 0-branes come from the D2-brane wrapped on the four non-geometric 2-cycles, the other four from the D4-brane wrapped on the four dual 4-cycles. The eight 1-branes arise from the D2-branes wrapped on the two 1-cycles, the D4-branes on the four 3-cycles and the D6-branes on the two 5-cycles. Similarly, one gets again more eight 2-branes and eight 3-branes.   A completely equivalent analysis can be carried out for the Type IIB theory, starting of course from the ten dimensional D$p$-branes with $p$ odd.  It should be stressed that we always get eight $\alpha=-1$ $p$-branes for any $p$, as reported in table \ref{branesIIBT6Z2Z2}, due to the fact that the wrapping rules for these branes never give any doubling.  In other words, one can think of $\alpha=-1$ branes as coming from {\it either} the Type IIA {\it or} the Type IIB theory in higher dimensions, and from {\it either} a geometric orbifold {\it or} a non-geometric one, getting always the same result.
\item $\alpha=-2$ branes. These branes always probe geometric cycles, and they double if they do not wrap. There is a NS5-brane in both the Type IIA and the Type IIB theory in ten dimensions. By wrapping a 4-cycle, it gives rise to a four dimensional 1-brane. The wrapping rules predict $3\times 2^2 =12$ such branes, where the $2^2$ comes from the two directions of the orbifold which the 5-brane does not wrap, in a precise agreement with the result in table \ref{branesIIBT6Z2Z2}. Similarly, the number of 2-branes is $8 \times 2^3 = 64$, again as in table \ref{branesIIBT6Z2Z2}, being $8$ the number of 3-cycles and $2^3$ the wrapping rules doubling. Finally, the number of 3-branes is $ 3 \times 2^4 \times \tfrac{1}{2} = 24$. The naive result has thus to be modified by the extra $1/2$ factor.  We have already discussed a similar phenomenon for the $\alpha=-2$ 5-branes that occur in the compactification on the orbifold $T^4 /\mathbb{Z}_N$ to six dimensions in the previous section, with the the extra factor $1/2$ arising because the branes split evenly between the  ${\cal N}=(1,1)$ and  ${\cal N}=(2,0)$ theories. This halving is essential to get the right counting also after the further dimensional reduction on $T^2$.  The  $T^6/(\mathbb{Z}_2 \times \mathbb{Z}_2 )$ orbifold can be thought of as a $\mathbb{Z}_2$ projection acting on the orbifold $( T^4/\mathbb{Z}_2 ) \times T^2$, and therefore the extra factor $1/2$ is inherited from the half-maximal theory.
\item $\alpha=-3$ branes.  In this case the wrapping rules always give a doubling, and since $\alpha$ is odd we have to consider both geometric and non-geometric cycles. The only $\alpha=-3$ brane present in ten dimensions is the 7-brane of the Type IIB theory, that is the S-dual of the D7-brane. When it wraps a geometric 6-cycle, the number of resulting 1-branes is $\tfrac{1}{4} \times 2^6 \times \tfrac{1}{2} = 8$, where the first factor is the relative weight of the geometric orbifold, the second comes from the wrapping rules and the last from the fact that the 7-brane is present only in the Type IIB theory. The result agrees with table \ref{branesIIBT6Z2Z2}. The $\tfrac{3}{4} \times 2 \times 2^6 \times \tfrac{1}{2} = 48$ 2-branes derive from the 7-brane wrapping non-geometric 5-cycles. Again, $\tfrac{3}{4}$ is the relative weight of non-geometric orbifolds and there are exactly $2$ non-geometric 5-cycles to reproduce the number present in the table. Finally, the 3-branes result from the 7-brane wrapping 4-cycles, which can either be geometric (three) or non-geometric (four). In the geometric case one gets $\tfrac{1}{4} \times 3 \times 2^6 \times \tfrac{1}{2} = 24$ branes, while in the non geometric case one gets $\tfrac{3}{4} \times 4 \times 2^6 \times \tfrac{1}{2} = 96$ branes. A comparison with the table shows that the 24 branes from the geometric cycles correspond $A_{4, M N_1 N_2 N_3}$, while the 96 branes from the non-geometric cycles are associated to the remaining 4-forms.
\item $\alpha=-4$ branes.  We limit ourselves to some comments on these branes.  As in the maximal and half-maximal case, indeed, we do not expect to derive all of them from the wrapping rules related to the $\alpha=-4$ 9-brane of Type IIB. This is obvious by looking at table \ref{branesIIBT6Z2Z2}, where the presence of 1-branes and 2-branes cannot be justified in terms of the 9-brane that, being space-filling, can only give rise to 3-branes. The wrapping rules give back $2^6 \times \tfrac{1}{2} = 32$ 3-branes, associated to the field $A_{4, M_1 M_2 A_1 A_2 ab}$. To be precise, decomposing the indices as done in eq. \eqref{fourdimdecomplightcone} for the $A_{2,M_1 M_2}$ field, and using eq. \eqref{alphaorbifoldT6Z2Z2}, one obtains the $\alpha=-4$ components
  \begin{eqnarray}
  & & A_{4 , i+ \  j-\  A_1 A_2 ab  } \quad \ \rightarrow \quad   2 \times 2^2 \times 2 =16 \ {\rm branes}\nonumber \\
  & & A_{4,  m_1 m_2 A_1 A_2 ab} \quad   \ \ \rightarrow \quad 2^2 \times 2^2 \times 2 =32 \ {\rm branes} \quad .
  \end{eqnarray}
Both components form an irreducible representation of the perturbative symmetry group, but the last correspond to the 32 branes predicted by the wrapping rules.  It should be stressed that, although the wrapping rules do not determine the full content in the $\alpha\le -4$, the branes related to reductions of the ten-dimensional ones fit exactly inside certain irreducible representations, analogously to what happens in more supersymmetric theories \cite{Bergshoeff:2012ex}.
\end{itemize}

\section{Comments on ${\cal N}=2$ string-string duality}\label{commduality}

There exist many ways to get theories with eight supercharges starting from compactifications of String/M/F-theory.  All of them should be equivalent, being connected by (perturbative or non-perturbative) string dualities.  In section \ref{wrultypeii} we have considered a low energy supergravity, assuming a massless spectrum and analyzing it in terms of a heterotic perspective as well as of a Type II perspective.  In a sense, assuming it as a Heterotic string compactified on ${\rm K3} \times T^2$ or as a Type II string compactified on a Calabi-Yau threefold, we have checked that the wrapping rules perfectly hold  and we have also made explicit the mapping among branes, dictated by the conjectured string-string duality.  Of course, while we have explicitly indicated the Type II model as a $\mathbb{Z}_2 \times \mathbb{Z}_2 $ orbifold of the Type IIA, the same is not equally evident from the heterotic perspective.  Actually, it is well known that Type II-Heterotic duality is very well established in six dimensions, where it is an S-duality between the Heterotic on $T^4$ and Type IIA on K3. It is relatively easy to exhibits checks on both sides of the duality, once one sits at a generic point in the moduli space where the non-abelian gauge group of the Heterotic string is broken to its maximal abelian subgroup.  There are also clear indications on how to deal with the point of enhanced non abelian symmetry in connection with K3-orbifold singularities, not visible in perturbation theory \cite{Aspinwall:1995zi}.

The situation is much more involved for what concerns the four dimensional duality connecting the Heterotic string on ${\rm K3} \times T^2$ with the Type IIA on a Calabi-Yau threefold, whose root seems to be, however, exactly the mentioned six-dimensional case.  First of all, the K3 manifold is basically unique while, on the contrary, there exist a very large number of classes of Calabi-Yau threefolds.  Of course, the difference is balanced, on the heterotic side, by the huge number of ways related to the choice of the gauge bundle with base the ${\rm K3} \times T^2$ manifold.  However, are exactly the properties of the gauge bundle together with  the presence of the $B_2$ field that make the duality quite subtle to check.
Indeed, due to the anomaly cancellation, the gauge bundle is not flat and must have instanton number equal to 24. Moreover, in six dimensions $n_H-n_V=244$, and a remnant of these numbers still survives in four dimensions: for instance, performing first a K3 compactification and then a further  reduction on $T^2$, the number of hypermultiplets does not change and at least three vector multiplets are expected, coming from the toroidal reduction of the heterotic geometric moduli.  Dual Type II models, in this case, require at least $h_{11} \ge 3$ on the  Type IIA side and $h_{12} \ge 3$ on the Type IIB side.  The gauge bundle does contribute both to the vector multiplet moduli space, a special K\"ahler manifold, and to the hypermultiplet moduli space, whose generic structure is actually more difficult to characterize \cite{Alexandrov:2014jua}.  Moreover, the duality is no longer an S-duality.  Indeed, the fact that the heterotic dilaton is part of a vector multiplet while the Type II dilaton is in a hypermultiplet, characterizes the quantum corrections.  For instance, the corrections to ${\cal{M}}_V$ in the Heterotic compactification come from target space instantons, while on the Type II side they come from world-sheet instantons \cite{Ferrara:1995yx}.  To compare the two vacua, one has to require that the Heterotic string be weakly coupled, namely the heterotic dilaton must approach $-\infty$ and, simultaneously, its dual field, the size of a holomorphic curve of the Calabi-Yau, must be very large.  In order for this phase to exists, the Calabi-Yau threefolds $X$ and its mirror $\tilde{X}$ on which the Type II are propagating  must be K3 fibrations over $\mathbb{P}^1$ \cite{Klemm:1995tj, Aspinwall:1995vk}, and the modulus dual to the dilaton is the size of the base manifold $\mathbb{P}^1$.  If also K3 is fibered over $\mathbb{P}^1$ (with fiber $T^2$), the duality can be interpreted as the six-dimensional one acting fiber-wise between K3 and $T^4$, at least when adiabatic arguments can be invoked \cite{Vafa:1995gm}. Starting from the seminal proposal in \cite{Kachru:1995wm, Ferrara:1995yx}, several examples of (chains of) dual pairs have been found (see, for instance, \cite{Aldazabal:1995yw, Vafa:1995gm, Candelas:1996su}).  The strategy consists typically in analyzing an Heterotic compactification by choosing the realization of ${\rm K3} \times T^2$ and the gauge bundle on it with instanton number 24.  Looking at the surviving gauge group (possibly after compatible Higgsing or breaking) and considering the phase where the group itself is reduced to the Cartan component, one gets the number $n_V$ of vector multiplets and the number $n_H$ of hypermultiplets.  Type II candidate dual pairs are then compactifications on Calabi-Yau owning the same Hodge numbers (excluding in the counting the universal hypermultiplet). The comparison of the moduli space on the two sides of the duality is indispensable to support the existence of the conjectured pairs.  Many checks regarding the vector moduli space have been realized, because of the special K\"ahler structure allowing a description of the low energy effective action in terms solely of the prepotential \cite{Aspinwall:1995vk}.  For the moduli in the hypermultiplet sector, there have been recently many progresses \cite{Louis:2011aa, Alexandrov:2013yva}, even though the map is less understood.  Other instances are related to M- or F-theory realizations of the duality.

A crucial point, however, is that the conjectured duality holds independently on the checks that can be done searching for dual pairs by matching the spectra, the moduli spaces and the quantum corrections, and also independently on the geometric realization of the four-dimensional string vacua.  Actually, some of the dual pairs are built using freely acting orbifold without a geometric interpretation.  In section \ref{z2xz2} our Type II (geometric and non-geometric) orbifold model is related to a compactification on a Calabi-Yau manifold with Hodge numbers $(51,3)$, and its mirror.  It falls within the class of K3 fibrations, as can be verified in \cite{Avram:1996pj}.  We did not search for the explicit geometric Heterotic dual, that would require a careful study of the gauge bundle and of the moduli potential in the low energy approximation, to check the existence of suitable flat direction that allow to properly break the gauge group to its maximal abelian subgroup, keeping the ${\cal N}=2$ supersymmetry unbroken.  Assuming the effective supergravity spectrum we checked, instead, the validity of the wrapping rules.  It comes from an average on the T-duality group orbit, merging both geometric and non-geometric compactifications. This fact  should be understood, in our opinion, as a clear indication that the duality is at work in the moduli space of String/M/F-theory, avoiding the construction of explicit geometric dual pairs where it can be verified.  In other words, the validity of the wrapping rules is a {\it necessary} condition once one assumes an underlying unique theory, being referred to the classification of single brane states rather than to (bound state) configurations.  The possibility of analyzing the whole orbit of the T-duality group clearly depend on the simplicity of the $\mathbb{Z}_2 \times \mathbb{Z}_2$ orbifold.  It would be obviously very interesting to extend the same control over other, more general, orbifolds, and in general over Calabi-Yau and flux compactifications, as well as to theories with less than eight supercharges.  We postpone the discussion of these issues to future work.

\section{Conclusions}

In this paper we have shown that the wrapping rules, satisfied by the branes of the ten-dimensional Type IIA and Type IIB string theories upon torus dimensional reduction, are also valid for the compactification on the orbifold $T^6/(\mathbb{Z}_2 \times \mathbb{Z}_2)$. A crucial ingredient is the observation that
the lower-dimensional theory must be considered not only as arising from  either Type IIA or Type IIB theories, but also from either geometric or non-geometric T-dual orbifolds. Only if this information is implemented correctly, the wrapping rules give the right number of branes. This generalizes the results of \cite{Bergshoeff:2013spa}, showing how the branes of the six-dimensional theory in the compactification of Type IIA on the orbifold $T^4/\mathbb{Z}_N$ satisfy the wrapping rules, provided that one also considers the same theory as arising from Type IIB on the non-geometric T-dual orbifold.

As already mentioned in the previous section, it would be interesting to generalize this procedure to other six-dimensional orbifolds in analogy with what was done in \cite{Bergshoeff:2013spa}, where the $T^4/\mathbb{Z}_2$ analysis of \cite{Bergshoeff:2012jb} was extended to any four-dimensional $\mathbb{Z}_N$-orbifold. The difference with respect to the four-dimensional case is that the resolution of the singularities of six-dimensional orbifolds leads to Calabi-Yau manifolds, whose topology is far from being unique and whose structure of the T-duality group orbit is very difficult to determine. We hope to report on these generalisations in the near future.

The consistency of the wrapping rules, verified in these simple settings both in heterotic and in Type II compactifications, is a clear indication of the 
validity of string-string dualities, independently of the explicit checks on dual paired theories.  Indeed, the classification of 1/2-BPS single $p$-brane states is universal and holds regardless of the compactification details.  In other words, assuming an underlying unique theory, the validity of the wrapping rules is a necessary condition, once taken correctly into account the T-duality group orbits.

In the paper we have derived for various theories, namely Type II theories compactified on $( T^4/\mathbb{Z}_2 ) \times T^n$ and on $T^6/(\mathbb{Z}_2 \times \mathbb{Z}_2 )$, the value of the dilaton weight $\alpha$ of each brane as a function of the rank and the number of specific lightlike internal indices of the corresponding field potential. In each theory, the relation between this value of $\alpha$ and the one that results in the Heterotic theory exploits the nature of the duality relating the two theories. In \cite{Bergshoeff:2014lxa} it was observed that the general brane classification in ${\cal N}=2$ theories, as well as in theories with sixteen supersymmetries (derived in \cite{Bergshoeff:2012jb}), allows to determine all the branes in the Heterotic theory that support vector multiplets. Using the definition of the value of $\alpha$ in the Heterotic theory and the one in the dual Type II theory, one can determine what are the values of $\alpha$ of the vector branes in both cases. While on the heterotic side the vector branes have $\alpha=-4$ or more negative \cite{Bergshoeff:2012jb}, in the Type II case the maximum value of the same $\alpha$ is $-1$ corresponding, as expected, to D-branes, end points of fundamental strings. This result can be appreciated by comparing tables \ref{braneshetK3T2} and \ref{branesIIBT6Z2Z2}. The vector branes correspond to the potential $A_{4, M N_1 N_2 N_3}$
leading, in the Heterotic theory, to 96 $\alpha=-4$ 3-branes.  They split, as shown in table \ref{branesIIBT6Z2Z2}, in Type II branes with different values of $\alpha$ including, in particular, 8 D3-branes.

To conclude, we would like to stress that the universality of the wrapping rules introduced in \cite{Bergshoeff:2011mh,Bergshoeff:2011ee} cannot be just considered as a numerical coincidence, but rather as an information on the different type of generalized geometry that each brane with different $\alpha$ probes. In determining the number of 1/2-supersymmetric branes, the geometry of the compactification space is only relevant in giving the number of various geometric supersymmetric cycles, where each brane has a universal behavior under reduction, depending only on the value of $\alpha$. This is still true in the case of non-geometric orbifolds arising from merging T-duality transformations with the orbifold group acting simultaneously on both $G$ and $B$.  However, in the non-geometric T-dual setting, the branes with  $\alpha=-1$ and $\alpha=-3$ probe {\it effective} cycles that are not geometric, in the sense that they are not cycles of the original orbifold \cite{Bergshoeff:2013spa}, but only of the T-dual one. Once these cycles are determined, the number of 1/2-supersymmetric branes again follows from the wrapping rules. It would be interesting to investigate how this can be understood in the context of double field theory \cite{Hohm:2013bwa}, where the background fields $G$ and $B$ are treated on the same footing from the start.

\section{Acknowledgments}

It is a pleasure to thank E.A. Bergshoeff, M. Petrini and Ya.S. Stanev for very interesting discussions, and A. Tomasiello for email correspondence.


\begin{thebibliography}{99}

\bibitem{Witten:1995ex}
  E.~Witten,
  ``String theory dynamics in various dimensions,''
  Nucl.\ Phys.\ B {\bf 443} (1995) 85
  [hep-th/9503124].

\bibitem{Hull:1994ys}
  C.~M.~Hull and P.~K.~Townsend,
  ``Unity of superstring dualities,''
  Nucl.\ Phys.\ B {\bf 438} (1995) 109
  [hep-th/9410167].

\bibitem{Polchinski:1998rq}
  For reviews see, {\it e.g.}, J.~Polchinski,
  ``String theory. Vol. 1: An introduction to the bosonic string,''
  Cambridge, UK: Univ. Pr. (1998) 402 p ;
  ``String theory. Vol. 2: Superstring theory and beyond,''
  Cambridge, UK: Univ. Pr. (1998) 531 p. .
  K.~Becker, M.~Becker and J.~H.~Schwarz,
  ``String theory and M-theory: A modern introduction,''
  Cambridge, UK: Cambridge Univ. Pr. (2007) 739 p. .

\bibitem{Townsend:1996xj}
  See, {\it e.g.}, P.~K.~Townsend,
  ``P-brane democracy,''
  In *Duff, M.J. (ed.): The world in eleven dimensions* 375-389
  [hep-th/9507048];
  P.~K.~Townsend,
  ``Four lectures on M theory,''
  In *Trieste 1996, High energy physics and cosmology* 385-438
  [hep-th/9612121].


\bibitem{Polchinski:1995mt}
  J.~Polchinski,
  ``Dirichlet Branes and Ramond-Ramond charges,''
  Phys.\ Rev.\ Lett.\  {\bf 75} (1995) 4724
  [hep-th/9510017].


\bibitem{Bergshoeff:2011qk}
  E.~A.~Bergshoeff and F.~Riccioni,
  ``The D-brane U-scan,''
  arXiv:1109.1725 [hep-th].

\bibitem{Kleinschmidt:2011vu}
  A.~Kleinschmidt,
  ``Counting supersymmetric branes,''
  JHEP {\bf 1110} (2011) 144
  [arXiv:1109.2025 [hep-th]].

\bibitem{Bergshoeff:2012ex}
  E.~A.~Bergshoeff, A.~Marrani and F.~Riccioni,
  ``Brane orbits,''
  Nucl.\ Phys.\ B {\bf 861} (2012) 104
  [arXiv:1201.5819 [hep-th]].

\bibitem{Bergshoeff:2005ac}
  E.~A.~Bergshoeff, M.~de Roo, S.~F.~Kerstan and F.~Riccioni,
  ``IIB supergravity revisited,''
  JHEP {\bf 0508} (2005) 098
  [hep-th/0506013];
    E.~A.~Bergshoeff, J.~Hartong, P.~S.~Howe, T.~Ortin and F.~Riccioni,
  ``IIA/IIB Supergravity and Ten-forms,''
  JHEP {\bf 1005} (2010) 061
  [arXiv:1004.1348 [hep-th]].

\bibitem{Bergshoeff:2006qw}
  E.~A.~Bergshoeff, M.~de Roo, S.~F.~Kerstan, T.~Ortin and F.~Riccioni,
  ``IIA ten-forms and the gauge algebras of maximal supergravity theories,''
  JHEP {\bf 0607} (2006) 018
  [hep-th/0602280].

\bibitem{Riccioni:2007au}
  F.~Riccioni and P.~C.~West,
  ``The E(11) origin of all maximal supergravities,''
  JHEP {\bf 0707} (2007) 063
  [arXiv:0705.0752 [hep-th]].

\bibitem{Bergshoeff:2007qi}
  E.~A.~Bergshoeff, I.~De Baetselier and T.~A.~Nutma,
  ``E(11) and the embedding tensor,''
  JHEP {\bf 0709} (2007) 047
  [arXiv:0705.1304 [hep-th]].

\bibitem{West:2001as}
  P.~C.~West,
  ``E(11) and M theory,''
  Class.\ Quant.\ Grav.\  {\bf 18} (2001) 4443
  [hep-th/0104081].

\bibitem{deWit:2008ta}
  B.~de Wit, H.~Nicolai and H.~Samtleben,
  ``Gauged Supergravities, Tensor Hierarchies, and M-Theory,''
  JHEP {\bf 0802} (2008) 044
  [arXiv:0801.1294 [hep-th]].

\bibitem{Nicolai:2000sc}
  H.~Nicolai and H.~Samtleben,
  ``Maximal gauged supergravity in three-dimensions,''
  Phys.\ Rev.\ Lett.\  {\bf 86} (2001) 1686
  [hep-th/0010076];
  B.~de Wit, H.~Samtleben and M.~Trigiante,
  ``On Lagrangians and gaugings of maximal supergravities,''
  Nucl.\ Phys.\ B {\bf 655} (2003) 93
  [hep-th/0212239].

\bibitem{Bergshoeff:2013sxa}
  E.~A.~Bergshoeff, F.~Riccioni and L.~Romano,
  ``Branes, Weights and Central Charges,''
  JHEP {\bf 1306} (2013) 019
  [arXiv:1303.0221 [hep-th]].

\bibitem{Bergshoeff:2011zk}
  E.~A.~Bergshoeff and F.~Riccioni,
  ``String Solitons and T-duality,''
  JHEP {\bf 1105} (2011) 131
  [arXiv:1102.0934 [hep-th]].

\bibitem{Bergshoeff:2011ee}
  E.~A.~Bergshoeff and F.~Riccioni,
  ``Branes and wrapping rules,''
  Phys.\ Lett.\ B {\bf 704} (2011) 367
  [arXiv:1108.5067 [hep-th]].


\bibitem{Bergshoeff:2011mh}
  E.~A.~Bergshoeff and F.~Riccioni,
  ``Dual doubled geometry,''
  Phys.\ Lett.\ B {\bf 702} (2011) 281
  [arXiv:1106.0212 [hep-th]].

\bibitem{Bergshoeff:2012jb}
  E.~A.~Bergshoeff and F.~Riccioni,
  ``Heterotic wrapping rules,''
  JHEP {\bf 1301} (2013) 005
  [arXiv:1210.1422 [hep-th]].


\bibitem{Bergshoeff:2013spa}
  E.~A.~Bergshoeff, C.~Condeescu, G.~Pradisi and F.~Riccioni,
  ``Heterotic-Type II duality and wrapping rules,''
  JHEP {\bf 1312} (2013) 057
  [arXiv:1311.3578 [hep-th], arXiv:1311.3578].

\bibitem{Bergshoeff:2014lxa}
  E.~A.~Bergshoeff, F.~Riccioni and L.~Romano,
  ``Towards a classification of branes in theories with eight supercharges,''
  JHEP {\bf 1405} (2014) 070
  [arXiv:1402.2557 [hep-th]].

\bibitem{Aspinwall:1996mn}
  See, {\it e.g.}, P.~S.~Aspinwall,
  ``K3 surfaces and string duality,''
  In *Yau, S.T. (ed.): Differential geometry inspired by string theory* 1-95
  [hep-th/9611137].

\bibitem{Kachru:1995wm}
  S.~Kachru and C.~Vafa,
  ``Exact results for N=2 compactifications of heterotic strings,''
  Nucl.\ Phys.\ B {\bf 450} (1995) 69
  [hep-th/9505105].

\bibitem{Ferrara:1995yx}
  S.~Ferrara, J.~A.~Harvey, A.~Strominger and C.~Vafa,
  ``Second quantized mirror symmetry,''
  Phys.\ Lett.\ B {\bf 361} (1995) 59
  [hep-th/9505162].

\bibitem{de Wit:1984px}
  B.~de Wit, P.~G.~Lauwers and A.~Van Proeyen,
  ``Lagrangians of N=2 Supergravity - Matter Systems,''
  Nucl.\ Phys.\ B {\bf 255} (1985) 569.
  

\bibitem{Aspinwall:2000fd}
  For a review see, {\it e.g.}, P.~S.~Aspinwall,
  ``Compactification, geometry and duality: N=2,''
  hep-th/0001001.


\bibitem{Vafa:1994rv}
  C.~Vafa and E.~Witten,
  ``On orbifolds with discrete torsion,''
  J.\ Geom.\ Phys.\  {\bf 15} (1995) 189
  [hep-th/9409188].

\bibitem{Angelantonj:2002ct}
  See, {\it e.g.}, C.~Angelantonj and A.~Sagnotti,
  ``Open strings,''
  Phys.\ Rept.\  {\bf 371} (2002) 1
   [Erratum-ibid.\  {\bf 376} (2003) 339]
  [hep-th/0204089].


\bibitem{Antoniadis:1999xk}
  I.~Antoniadis, E.~Dudas and A.~Sagnotti,
  ``Brane supersymmetry breaking,''
  Phys.\ Lett.\ B {\bf 464} (1999) 38
  [hep-th/9908023].


\bibitem{Strominger:1996it}
  A.~Strominger, S.~-T.~Yau and E.~Zaslow,
  ``Mirror symmetry is T duality,''
  Nucl.\ Phys.\ B {\bf 479} (1996) 243
  [hep-th/9606040].

  
\bibitem{Aspinwall:1995zi}
  P.~S.~Aspinwall,
  ``Enhanced gauge symmetries and K3 surfaces,''
  Phys.\ Lett.\ B {\bf 357} (1995) 329
  [hep-th/9507012].


\bibitem{Alexandrov:2014jua}
  S.~Alexandrov, J.~Louis, B.~Pioline and R.~Valandro,
  ``N=2 Heterotic-Type II duality and bundle moduli,''
  arXiv:1405.4792 [hep-th].

\bibitem{Klemm:1995tj}
  A.~Klemm, W.~Lerche and P.~Mayr,
  ``K3 Fibrations and heterotic type II string duality,''
  Phys.\ Lett.\ B {\bf 357} (1995) 313
  [hep-th/9506112].


\bibitem{Aspinwall:1995vk}
  P.~S.~Aspinwall and J.~Louis,
  ``On the ubiquity of K3 fibrations in string duality,''
  Phys.\ Lett.\ B {\bf 369} (1996) 233
  [hep-th/9510234].


\bibitem{Vafa:1995gm}
  C.~Vafa and E.~Witten,
  ``Dual string pairs with N=1 and N=2 supersymmetry in four-dimensions,''
  Nucl.\ Phys.\ Proc.\ Suppl.\  {\bf 46} (1996) 225
  [hep-th/9507050].

\bibitem{Aldazabal:1995yw}
  G.~Aldazabal, A.~Font, L.~E.~Ibanez and F.~Quevedo,
  ``Chains of N=2, D = 4 heterotic type II duals,''
  Nucl.\ Phys.\ B {\bf 461} (1996) 85
  [hep-th/9510093].

\bibitem{Candelas:1996su}
  P.~Candelas and A.~Font,
  ``Duality between the webs of heterotic and type II vacua,''
  Nucl.\ Phys.\ B {\bf 511} (1998) 295
  [hep-th/9603170].


\bibitem{Louis:2011aa}
  J.~Louis and R.~Valandro,
  ``Heterotic-Type II Duality in the Hypermultiplet Sector,''
  JHEP {\bf 1205} (2012) 016
  [arXiv:1112.3566 [hep-th]].

\bibitem{Alexandrov:2013yva}
  S.~Alexandrov, J.~Manschot, D.~Persson and B.~Pioline,
  ``Quantum hypermultiplet moduli spaces in N=2 string vacua: a review,''
  arXiv:1304.0766 [hep-th].


\bibitem{Avram:1996pj}
  A.~C.~Avram, M.~Kreuzer, M.~Mandelberg and H.~Skarke,
  ``Searching for K3 fibrations,''
  Nucl.\ Phys.\ B {\bf 494} (1997) 567
  [hep-th/9610154].


\bibitem{Hohm:2013bwa}
For a review see, {\it e. g.},  O.~Hohm, D.~L\"ust and B.~Zwiebach,
  ``The Spacetime of Double Field Theory: Review, Remarks, and Outlook,''
  Fortsch.\ Phys.\  {\bf 61} (2013) 926
  [arXiv:1309.2977 [hep-th]].



\end{thebibliography}
\end{document}